\documentclass[preprint]{imsart}

\usepackage{amsthm,amsmath,natbib}
\RequirePackage[colorlinks,citecolor=blue,urlcolor=blue]{hyperref}

\startlocaldefs

\usepackage[usenames, dvipsnames]{color}
\definecolor{mypurple}{RGB}{203, 66, 244}
\definecolor{mygray}{RGB}{145,145,145}
\definecolor{mygreen}{RGB}{0,128,0}

\usepackage{dsfont}

\usepackage{tikz}
\usetikzlibrary{positioning,shapes.geometric}

\usepackage{subcaption} 

\usepackage{geometry}
\usepackage{pdflscape}
\usepackage{afterpage}

\usepackage{booktabs} 

\usepackage{multirow}

\newcommand{\phm}{\phantom{-}}

\newcommand{\var}{\mbox{Var}}

\newcommand{\corr}{\mbox{Corr}}

\newcommand{\bA}{\bar{A}}

\newcommand{\bX}{\bar{X}}

\newcommand{\bY}{\bar{Y}}

\newcommand{\cL}{\mathcal{L}}

\def\ind{{\mathds 1}}

\endlocaldefs

\begin{document}

\begin{frontmatter}

\title{Linear mixed models with endogenous covariates: modeling sequential treatment effects with application to a mobile health study}
\runtitle{Linear mixed models under endogeneity}

\begin{aug}
\author{\fnms{Tianchen} \snm{Qian}\ead[label=e1]{qiantianchen@fas.harvard.edu}},
\author{\fnms{Predrag} \snm{Klasnja}\ead[label=e2]{
klasnja@umich.edu}}
\and
\author{\fnms{Susan A.} \snm{Murphy}\ead[label=e3]{samurphy@fas.harvard.edu}}
\address{Department of Statistics, Harvard University, Cambridge, MA 02138 \\ \printead{e1,e3}.}
\address{School of Information, University of Michigan, Ann Arbor, MI 48109 \\ \printead{e2}.}
\end{aug}


\runauthor{T. Qian, P. Klasnja and S. Murphy}

\bigskip

%
%
%


\begin{abstract}
Mobile health is a rapidly developing field in which behavioral treatments are delivered to individuals via wearables or smartphones to facilitate health-related behavior change. Micro-randomized trials (MRT) are an experimental design for developing mobile health interventions. In an MRT the treatments are randomized numerous times for each individual over course of the trial. Along with assessing treatment effects, behavioral scientists aim to understand between-person heterogeneity in the treatment effect. A natural approach is the familiar linear mixed model. However, directly applying linear mixed models is problematic because potential moderators of the treatment effect are frequently endogenous---that is, may depend on prior treatment. We discuss model interpretation and biases that arise in the absence of additional assumptions when endogenous covariates are included in a linear mixed model. In particular, when there are endogenous covariates, the coefficients no longer have the customary marginal interpretation. However, these coefficients still have a conditional-on-the-random-effect interpretation. We provide an additional assumption that, if true, allows scientists to use standard software to fit linear mixed model with endogenous covariates, and person-specific predictions of effects can be provided. As an illustration, we assess the effect of activity suggestion in the HeartSteps MRT and analyze the between-person treatment effect heterogeneity.
\end{abstract}



\end{frontmatter}

\section{Introduction}

Mobile health (mHealth) refers to the use of mobile phones and other wireless devices to improve health outcomes, often by providing individuals with support for health-related behavior change. One major category of time-varying treatments delivered through mobile devices, which is the focus of this paper, are ``push interventions''; in this setting, the mobile device determines when a treatment will be provided, rather than the individual seeking the intervention of her own accord (e.g., by opening the app). Push interventions are usually provided via some kind of a notification, such as an audible ping, vibration, or the lock screen of a phone lightening up. For example, to encourage physical activity in sedentary individuals, the HeartSteps intervention sends users push notifications that contain contextually-tailored activity suggestions \citep{klasnja2018}. 

Micro-randomized trials (MRTs) provide an experimental design for developing mHealth interventions. These trials provide longitudinal data to assess whether there is an effect of a time-varying treatment, how this effect changes over time, and whether aspects of the current context impact the effect \citep{liao2016sample, dempsey2015randomised}. In an MRT, each individual is randomized repeatedly to different versions of a treatment (or no treatment) with a known probability over the course of the trial (often hundreds or even thousands of times). Between randomizations, the trial collects covariate data on the individual's current/recent context via sensors and self-report, and after each randomization it assesses a proximal outcome. The large number of randomization points likely covers a wide range of contexts, and methods that exploit this for assessing effect moderation of a time-varying treatment have been developed \citep{boruvka2017}.

Random effects models \citep{laird1982random,raudenbush2002}, sometimes also known as mixed effect models, hierarchical models, or multilevel models, have been used with great success in the analysis of longitudinal studies. Behavioral scientists, and researchers from many other scientific fields, have long used random effects model in research involving longitudinal data \citep{agresti2000,berger2004robust,cheung2008model,luger2014robust}. A particularly appealing feature of random effects models is the ability to predict person-specific random effects, which enables quantitative characterization of between-person heterogeneity due to unobserved factors \citep{schwartz2007analysis,bolger2013intensive}. 
Understanding such heterogeneity can bring forth new scientific hypotheses for further studies. In addition, the random effects provide a model for the within-person dependence in the time-varying outcome, which improves efficiency in parameter estimation. Because data from an MRT is longitudinal, it is natural to consider a random effects model when making inference about treatment effects using MRT data.

However, random effects models were designed for settings where the covariates are considered fixed, and inferential challenges arise when one tries to apply the standard random effects model if there are endogenous time-varying covariates. A time-varying covariate is endogenous if this covariate is not independent of previous treatment or outcomes; we give a more precise definition in Section \ref{subsec:notation}. As written above, MRTs are conducted to make inference about the effect of a time-varying treatment, how this effect changes over time, and whether certain aspects of the current context impact the effect. Covariates, often endogenous, describe the individual's context, and it is often of scientific interest to assess if the time-varying treatment is moderated by certain endogenous covariates. Furthermore, to reduce variance in assessing treatment effects, it is very useful to control for an endogenous covariate in the analysis \citep{boruvka2017}.
For example, consider HeartSteps, an MRT of an intervention that aims to increase physical activity among sedentary adults \citep{klasnja2018}. In this study the treatments are contextually-tailored activity suggestions. The steps taken by the individual during the 30 minutes prior to randomization is likely highly correlated with the primary proximal outcome, the step count in the subsequent 30 minutes. Thus it is useful to control for this covariate in the analysis as well as to assess whether this covariate moderates the effect of the activity suggestion on the subsequent 30-minute step count. However, because the activity suggestions are randomized roughly every 2 hours, it is likely that the 30-minute step count prior to randomization is related to past step counts (i.e., past outcomes) as well as past treatment, which makes it an endogenous covariate. As we discuss below, including endogenous covariates in random effects models can result in biased estimates.

A related but different concept to an endogenous covariate is a time-varying confounder. Recall that a time-varying confounder, sometimes also called a time-dependent confounder, is a covariate that is affected by previous treatment (hence is endogenous) and affects future treatment assignment \citep{daniel2013methods, hernan2019causal}. To our surprise, even without time-varying confounding (e.g., when the randomization probability is constant in an MRT), the inclusion of endogenous covariates  in random effects models can cause bias in assessment of the treatment effects.

\citet{pepeanderson1994} pointed out that when using generalized estimating equations (GEE) with endogenous covariates, one should use working independence correlation structure to avoid biased estimates.  \citet{diggle2002}, in their classic monograph on longitudinal data analysis, noted that:

\begin{quote}
``Although \citet{pepeanderson1994} focused on the use of GEE, the issue that they raise is important for all longitudinal data analysis methods including likelihood-based methods such as linear and generalized linear mixed models.''
\end{quote}
In this paper, we focus on linear mixed models (LMM), a simple form of random effects models where the outcome is continuous and the link function is identity. We review how problems arise when endogenous covariates are included in LMM. Coefficients, and specifically treatment effects, in a standard LMM with fixed covariates have both marginal and conditional-on-the-random-effect interpretations. But the marginal interpretation is no longer valid with endogenous covariates. 

Fortunately, despite losing the marginal interpretation, the conditional interpretation of the parameters is consistent with scientific interest in the prediction of person-specific effects in MRTs.  Here we propose to  interpret treatment effects as conditional on the random effect in LMM with possibly endogenous covariates. We provide an additional assumption under which valid estimates of the effect (conditional on the random effect) of the time-varying treatment, estimates of the variance components, and person-specific predictions of these treatment effects can be obtained through standard LMM software, even if some covariates are endogenous. Simulation studies are conducted to support the main result.

Lastly, we discuss whether and when the aforementioned assumption makes sense in HeartSteps, and analyze the data using the proposed method.

The paper is organized as follows. We provide an overview of the HeartSteps MRT in Section \ref{subsec:motivating-example}. We introduce notation and definition in Section \ref{subsec:notation}. In Section \ref{sec:explain-issue} we give a detailed account of the issue regarding endogenous covariates in a standard LMM, and review related literature in causal inference (Section \ref{subsec:time-varying-confounding}) and econometrics (Section \ref{subsec:econometric}). Next we provide an assumption under which treatment effects can be estimated based on LMM with endogenous covariates in Section \ref{sec:model}. 
In Section \ref{sec:simulation} we present results from a simulation study. We apply the proposed model to analyzing the HeartSteps data in Section \ref{sec:data-analysis}. Section \ref{sec:discussion} concludes with discussion.

\subsection{Motivating Example: HeartSteps}
\label{subsec:motivating-example}

Our motivating example is from HeartSteps, a 6-week MRT of an mHealth intervention to encourage regular walking among sedentary adults \citep{klasnja2018}. The intervention package in HeartSteps includes multiple components; in this paper we focus on one push intervention component as the treatment, which is the activity suggestions. Each individual is in the study for 42 days, and is randomized 5 times a day, each time with probability 0.6 to receive an activity suggestion. The 5 randomization times are pre-specified and individual-specific, corresponding to each individual's morning commute, lunchtime, mid-afternoon, evening commute, and after-dinner. The content of the suggestion was tailored to the current time of day, weekend vs weekday, weather, and the individual’s current location. The activity suggestions were designed to help individuals get activity throughout the day. Due to the tailoring of the suggestions to the individual’s current context, the research team expected to see the greatest impact of the activity suggestions on near time, proximal activity, so the proximal outcome is defined as the individual's step count during the 30 minutes following each randomization. In addition to the step counts, at each randomization the individual's context is also recorded, including current location, weather and 30-minute step count prior to randomization. Note that the 30-minute step count prior to the time of randomization is likely impacted by prior treatment and thus is an endogenous covariate. In addition to the measured information, there are other unobserved variables that may impact the treatment effect, such as each individual's commitment to becoming more active, conscientiousness, degree of social support and so on. Therefore, it is of interest to provide person-specific predictions of treatment effect. We will apply methods developed in this paper to the HeartSteps data in Section \ref{sec:data-analysis}.

\subsection{Notation and definition}
\label{subsec:notation}

We will consider two settings in the paper. In the first setting we consider a longitudinal study without treatment, and in the second one with a sequentially randomized treatment. The first setting will be used to explain bias incurred by the inclusion of endogenous covariates in random effects models, as this issue also occurs without treatment and is easier to explain there. The second setting involves time-varying treatment that is sequentially randomized; thus it's relevant to data from MRTs. We will see that randomized treatment assignment in MRT does not necessarily alleviate the biases resulting from the inclusion of endogenous time-varying covariates in LMMs. We will consider assumptions that allow valid estimation under this second setting. The setting under consideration will be clear from the context.

For the first setting without treatment, we denote data for individual $i$ by $X_{i1}
, Y_{i2}$, $X_{i2}, Y_{i3}, \ldots, X_{iT_i},$ $Y_{iT_i+1}$, where $T_i$ denotes the total number of observations for individual $i$. $X_{it}$ is a vector of covariates prior to the $t$-th time point and $Y_{it+1}$ is the outcome subsequent to the $t$-th time point. Note that the time index for the outcome $Y$ is augmented by $1$ to make it consistent with the second setting. We use overbar to denote history; for example, $\bX_{it} = (X_{i1}, X_{i2}, \ldots, X_{it})$. The individual's history information up to the $t$-th time is denoted by $H_{it} = (X_{i1}, Y_{i2}, \ldots, X_{it-1}, Y_{it}, X_{it}) = (\bY_{it}, \bX_{it})$.

For the second setting with treatment, the data for individual $i$ is $X_{i1}, A_{i1}
, Y_{i2}, X_{i2}$, $A_{i2}, Y_{i3}, \ldots, X_{iT_i},$ $A_{iT_i},Y_{iT_i+1}$, where $X_{it}$ is the covariate vector prior to the $t$-th time, $A_{it}$ is the randomized treatment at the $t$-th time, and $Y_{it+1}$ is the proximal outcome subsequent to the $t$-th time. To maintain expositional clarity, throughout we assume there are only two types of treatment and $A_{it}\in \{0,1\}$. The history is defined as $H_{it} = (X_{i1}, A_{i1}, Y_{i2}, \ldots, X_{it-1}, A_{it-1}, Y_{it}, X_{it}) = (\bY_{it}, \bX_{it}, \bA_{it-1})$. We define $X_{i0} = \emptyset$, $A_{i0} = \emptyset$, and $Y_{i1} = \emptyset$.

In both settings, we use $b_i$ to denote the random effect of individual $i$. 


We use $\perp$ to denote statistical independence; for example, $A \perp B \mid C$ means that $A$ is independent of $B$ conditional on $C$. In the first setting, a covariate process $X_{it}$ is called \textit{exogenous} (with respect to the outcome process $Y_{it}$) if $X_{it} \perp \bY_{it} \mid \bX_{it-1}$; otherwise, $X_{it}$ is \textit{endogenous}. In the second setting, $X_{it}$ is called \textit{exogenous} if $X_{it} \perp (\bY_{it}, \bA_{it-1}) \mid \bX_{it-1}$; otherwise, $X_{it}$ is \textit{endogenous}. In a longitudinal study, examples of exogenous covariates include baseline variables (age, gender, etc.), functions of time, and time-varying variables that are not impacted by prior treatment or prior outcome, such as weather. 



\section{Issue of linear mixed models with endogenous covariates}
\label{sec:explain-issue}

In this section, we start by considering the situation where no treatment is involved, as endogenous covariates give rise to issues even without considering causal inference. We give a brief review of standard LMM in Section \ref{subsec:review-lmm}, and explain the issue of endogenous covariates in Section \ref{subsec:explain-issue}. In Section \ref{subsec:time-varying-confounding}, we briefly review causal inference literature on a related topic, time-varying confounding, which is a more restrictive definition than endogeneity. In Section \ref{subsec:econometric}, we discuss connections to the econometric literature. We comment on why the methods reviewed in Sections \ref{subsec:time-varying-confounding} and \ref{subsec:econometric} do not directly solve the issue of LMM with endogenous covariates in MRTs.

\subsection{Brief overview of standard LMM with exogenous covariates}
\label{subsec:review-lmm}

A standard linear mixed model (LMM) \citep{laird1982random} assumes a relationship between the covariate $X_{it}$ and the outcome $Y_{it+1}$ such as the following:
\begin{equation}
Y_{it+1} = X_{it}^T \beta + Z_{it}^T b_i + \epsilon_{it+1}. \label{eq:standard-lmm}
\end{equation}
Here, $b_i \sim N(0,G)$ denotes the vector of person-specific random effects, $Z_{it} \subset X_{it}$
and $\epsilon_{it+1} \sim N(0,\sigma^2_\epsilon)$ is a random noise. It is typically assumed that $\epsilon_{it+1}$'s are independent of each other and of $b_i$, and we will adopt this assumption throughout this paper. 
This model specifies the conditional distribution of $Y_{it+1}$ given $X_{it}$ and $b_i$; in particular, this is a Gaussian distribution with mean:
\begin{equation}
E(Y_{it+1} \mid X_{it}, b_i) = X_{it}^T \beta + Z_{it}^T b_i. \label{eq:lmm-conditional-model}
\end{equation}
Furthermore, use of the standard LMM assumes, though not always explicitly, that all covariates are fixed, or at least exogenous and independent of $b_i$. Thus, the marginal mean of $Y_{it}$ is
\begin{equation}
E(Y_{it+1} \mid X_{it}) = X_{it}^T \beta, \label{eq:lmm-marginal-model}
\end{equation}
because $E(b_i \mid X_{it})=0$.
Thus, when the covariates are exogenous and independent of $b_i$, $\beta$ has both a conditional interpretation and a marginal interpretation\footnote{In this paper, we use the term ``conditional (model/interpretation)'' to denote a model that is conditional on the random effect, and we use ``marginal (model/interpretation)'' to denote a model where the random effect is marginalized over. This is consistent with the terminology in \citet{zeger1992overview} and \citet{heagerty2000marginalized}.}. This dual interpretation provides the opportunity to estimate $\beta$ with alternative approaches such as with generalized estimating equations (GEE) \citep{zeger1986}, depending on the desired robustness of the estimator of $\beta$ to deviations from the LMM assumptions. 

Assuming the covariates are indeed exogenous and independent of $b_i$, the maximum likelihood score equation for $\beta$ is:
\begin{align}
\frac{1}{n} \sum_{i=1}^n X_i V_i^{-1} (Y_i - X_i^T \beta) = 0, \label{eq:lmm-ee}
\end{align}
where $X_i = (X_{i1}, \ldots, X_{iT_i})$, $Z_i = (Z_{i1}, \ldots, Z_{iT_i})$ and $Y_i = (Y_{i2}, \ldots, Y_{iT_i+1})^T$, $V_i = Z_i^T G Z_i + R_i$ is a $T_i \times T_i$ covariance matrix, and $R_i$ is a $T_i \times T_i$ diagonal matrix with all diagonal entries equal to $\sigma^2_\epsilon$.

\subsection{Issue with endogenous covariates: marginal interpretation is no longer valid}

\label{subsec:explain-issue}

Any LMM solves the same estimating equation as a GEE with a corresponding non-independence working correlation structure (e.g., an LMM with a random intercept solves the same estimating equation as a GEE with compound symmetric working correlation structure). In fact, \eqref{eq:lmm-ee} is the estimating equation for GEE with marginal mean model \eqref{eq:lmm-marginal-model} and working correlation matrix $V_i$. In the GEE literature, estimation bias due to the inclusion of endogenous covariates has been discussed repeatedly. We first review this briefly.

\citet{pepeanderson1994} first pointed out that when using GEE to estimate parameters in $E(Y_{it+1} \mid X_{it})$, a sufficient condition for estimation consistency is either
\begin{equation}
E(Y_{it+1} \mid X_{it}) = E(Y_{it+1} \mid X_{i1}, \ldots, X_{iT}) \label{eq:FCCM}
\end{equation}
or the use of a working independence correlation structure. When \eqref{eq:FCCM} is violated and a correlation structure other than working independence is used, they provided simulation results to show that bias could occur. \citet[Chapter 12]{diggle2002} reiterated this point, and referred to \eqref{eq:FCCM} as ``full covariate conditional mean (FCCM)'' assumption. \citet{schildcrout2005} analyzed the bias-efficiency trade-off associated with working correlation choices of GEE for longitudinal binary data, when FCCM is violated due to exogenous covariates being time-varying, through simulation studies. This potential bias from the violation of FCCM have also been warned about by \citet{pan2000} in the context of linear regression via analytic calculations. \citet{tchetgen2012specifying} showed, in the context of marginal structural models \citep{robins1998MSM}, that when GEE is combined with inverse probability weighting for handling dropout, parameter estimation is generally biased in the presence of endogenous covariates unless either a condition similar to \eqref{eq:FCCM} holds or a working independence correlation structure is used.

When there are endogenous covariates, the FCCM assumption \eqref{eq:FCCM} is unlikely to hold because $Y_{it+1}$ may impact future $X_{is}$ for $s \geq t+1$. In this case, \citet{pepeanderson1994} suggested the use of working independence GEE to guarantee consistent estimation of parameters in $E(Y_{it+1} \mid X_{it})$.
Because of the close tie between the estimating equations of LMM and GEE, Pepe and Anderson's point about GEE implies that estimators fitted using the standard LMM could be inconsistent when there are endogenous covariates. Indeed, if one intends to estimate parameters in the marginal mean $E(Y_{it+1} \mid X_{it})$, then using LMM \textit{as an estimation procedure} can result in inconsistent estimators because of the biased estimating equations. However, in our opinion, this is not the fundamental issue of LMM under endogeneity, but rather a technical consequence.

More fundamentally, when there are endogenous covariates, LMM \eqref{eq:standard-lmm} \textit{as a model} can imply a marginal mean relationship different from \eqref{eq:lmm-marginal-model}. $X_{it}$ being endogenous means it may depend on previous outcomes, which in turn implies dependence on the random effect $b_i$. Thus, $E(b_i \mid X_{it})$ is usually nonzero and the conditional model \eqref{eq:lmm-conditional-model} may no longer imply the marginal model \eqref{eq:lmm-marginal-model}. The marginal model implied by \eqref{eq:lmm-conditional-model} becomes, instead,
\begin{equation}
E(Y_{it+1} \mid X_{it}) = X_{it}^T \beta + Z_{it}^T E(b_i \mid X_{it}). \label{eq:lmm-marginal-model-endogenous}
\end{equation}
As a concrete example, consider the case where each individual is observed for 2 time points ($T_i=2$), and the covariate at the second time point is the lag-1 outcome: $X_{i2} = Y_{i2}$. Suppose the variables are generated from the following LMM with a random intercept: $b_i\sim N(0,\sigma^2_u)$, $X_{i1} \sim N(0,\sigma^2_{X_1})$ independently of $b_i$, $Y_{i2} \mid X_{i1}, b_i \sim N(\beta_0 + \beta_1 X_{i1} + b_i, \sigma^2_\epsilon)$, $X_{i2} = Y_{i2}$, and $Y_{i3} \mid X_{i1}, Y_{i2}, X_{i2}, b_i \sim N(\beta_0 + \beta_1 X_{i2} + b_i, \sigma^2_\epsilon)$. This implies a parsimonious conditional relationship: $E(Y_{it+1} \mid X_{it}, b_i) = \beta_0 + \beta_1 X_{it} + b_i$,
but the induced marginal relationship is rather complex:
\begin{align}
E(Y_{i2} \mid X_{i1}) & = \beta_0 + \beta_1 X_{i1}, \nonumber \\
E(Y_{i3} \mid X_{i2}) & = ( 1- \rho \zeta - \rho ) \beta_{0} + \{ (1- \rho \zeta) \beta_{1}+\rho \} X_{i2}, \nonumber
\end{align}
with $\rho = \sigma^2_u / (\sigma^2_u + \sigma^2_\epsilon)$ and $\zeta = \beta_{1} \sigma_{X_{1}}^{2} / (\beta_{1} \sigma_{X_{1}}^{2}+\sigma_{u}^{2}+\sigma_{\epsilon}^{2})$.

\smallskip

Therefore, when building LMM with endogenous covariates, one needs to be aware that the modeling assumption is on the conditional relationship $E(Y_{it+1} | X_{it}, b_i)$, not the marginal relationship $E(Y_{it+1} | X_{it})$. Although it is attractive to treat $\beta$ in \eqref{eq:standard-lmm} with not only a conditional interpretation but also a marginal interpretation, which is true with exogenous covariates, the latter interpretation can be invalid with endogenous covariates. In addition to this model interpretation issue, endogenous covariates also give rise to additional concerns in model fitting, which will be discussed in Section \ref{sec:model}.

As a side note, for generalized linear mixed models, it is well known that even when all covariates are exogenous, the conditional parameter and the marginal parameter are different due to the nonlinear link function, and there has been work in the literature on connecting the two interpretations \citep{zeger1988models, heagerty1999marginally, wanglouis2004}. For LMMs, the discrepancy in the two interpretations  only occurs when there are endogenous covariates.

\subsection{Connection to time-varying confounding in causal inference literature}

\label{subsec:time-varying-confounding}

In the setting with treatment, a related issue, often called ``time-varying confounding'' or ``time-dependent confounding'', has been well studied in the causal inference literature. A time-varying covariate is a time-varying confounder if it is affected by previous treatment (hence is endogenous) and it affects future treatment assignment \citep{daniel2013methods, hernan2019causal}. Time-varying confounders are usually intermediate variables (that lie in the causal pathway between the treatment and the outcome), and this gives rise to inferential challenges for conventional regression-based methods due to the following dilemma: confounders should be adjusted for in the analysis, but intermediate variables should not \citep{diggle2002}.

Causal inference methods have been developed to estimate treatment effects in the presence of time-varying confounding. These methods include g-computation \citep{robins1986new}, structural nested models \citep{robins1994snmm, robins1997snmm}, inverse probability weighting in marginal structural models \citep{robins1998MSM, robins2000marginal}, history-restricted marginal structural models \citep{neugebauer2007causal}, sequential conditional mean models \citep{vansteelandt2007confounding, keogh2017analysis}, and weighted and centered least-squares for MRTs \citep{boruvka2017}. These methods cover a variety of estimands that characterize the effect of a time-varying treatment from various aspects, but all the treatment effects are marginal in the sense that no random effect is considered.

Estimators of conditional-on-the-random-effect versions of the above estimands will be potentially biased as  discussed in Section \ref{subsec:explain-issue}. Furthermore, the issue with bias persists even when $A_{it}$ is not confounded by observed or unobserved variables (e.g., when the randomization probability is constant). Take, for example, the sequential conditional mean models in \citet{vansteelandt2007confounding}, which considers the marginal expected mean $E(Y_{it+1} \mid \bA_{it}, \bX_{it})$. When random effect is incorporated, the model becomes the conditional expected mean $E(Y_{it+1} \mid \bA_{it}, \bX_{it}, b_i)$. When $X_{it}$ is endogenous, even if $X_{it}$ does not confound $A_{it}$, the same argument in Section \ref{subsec:explain-issue} applies, and the parameter in the conditional model $E(Y_{it+1} \mid \bA_{it}, \bX_{it}, b_i)$ generally does not have the marginal interpretation. This means the methods for estimating marginal treatment effect cannot be used to estimate parameters in the conditional model, let alone used to predict the random effects in the conditional model.

\subsection{Connection to level-2 endogeneity in econometric literature}

\label{subsec:econometric}

Violation of the assumption that the random effect being independent of the covariates, $b_{i} \perp X_{it}$, is sometimes called ``level-2 endogeneity'' in the econometric literature \citep{wooldridge2002econometric, grilli2011role}. It is well known that level-2 endogeneity can lead to biased parameter estimates \citep{ebbes2004regressor}; in particular, \citet{kim2007multilevel} gave a display similar to \eqref{eq:lmm-marginal-model-endogenous}, and warned about the bias that could occur when one uses an estimator intended for the marginal parameter (such as the ordinary least-squares) to estimate the conditional parameter---this is the counterpart of our discussion in Section \ref{subsec:explain-issue}, that using LMM to estimate the marginal parameter will incur bias with endogenous covariates.

Various estimators have been proposed in the econometric literature for the conditional parameter under level-2 endogeneity, many of which are based on explicitly modeling the conditional distribution of the random effects given the endogenous covariates \citep{mundlak1978pooling}, centering the time-varying covariate and the time-varying outcome by their average over time \citep{hausman1981panel, arellano1995another, neuhaus2006separating, kim2006omitted, hanchane2012solving}, constructing internal instrumental variables \citep{amemiya1986instrumental, arellano1991some, semykina2010estimating}, or using semiparametric efficiency theory by not specifying the distribution of the random effects \citep{liu2014semiparametric, garcia2016optimal}.

In those works, it is usually assumed that the error term $\epsilon_{it}$ is independent of the  history of the time-varying covariate, $\bX_{iT_i}$; thus these methods are not directly applicable to the MRT setting where future covariates can depend on previous outcomes (hence previous error terms). In addition, many of these methods focus on estimating the conditional parameter while treating the random effect as a nuisance parameter. We argue that in MRTs, prediction of the random effects are of equal importance to estimation of the conditional parameter; otherwise, one could have used the causal inference methods mentioned in Section \ref{subsec:time-varying-confounding} to estimate the marginal treatment effect. It is an open question whether the ideas behind the above methods can be adapted for LMM-based inference in MRTs.

\section{A conditional independence assumption}
\label{sec:model}


In an MRT, the observed history up to time $t$ is defined as $H_{it} = (X_{i1}, A_{i1}, Y_{i2}, \ldots, X_{it-1}, A_{it-1}, Y_{it}, X_{it})$. We consider the following LMM:
\begin{equation}
Y_{it+1} = f_0(H_{it})^T \beta_0 + A_{it} f_1(H_{it})^T \beta_1 + g_0(H_{it})^T b_{0i} + A_{it} g_1(H_{it})^T b_{1i} + \epsilon_{it+1}
\label{eq:proposed-model}
\end{equation}
for $t = 1, \ldots, T$, where $f_0(H_{it}), f_1(H_{it}), g_0(H_{it}), g_1(H_{it})$ are known functions of $H_{it}$.
For example, if we believe that the outcome depends linearly on time, current covariate and previous outcome, that the treatment also interacts with these three variables, and that the outcome has no residual association with other information in $H_{it}$, we may set each of $f_0(H_{it}), f_1(H_{it}), g_0(H_{it}), g_1(H_{it})$ to be $(1, t, X_{it}, Y_{it})$.
Recall that for simplicity we consider only binary treatment. In this section, we provide an additional assumption that, if true, ensures valid treatment inference and person-specific predictions via standard software even when there are endogenous covariates.

We make the standard LMM assumptions. The random effects $(b_{0i}^T, b_{1i}^T)$ are assumed to marginally follow a multivariate Gaussian distribution with mean 0 and variance-covariance matrix $G$. $A_{it}$ is assumed to be randomized with randomization probability depending only on $H_{it}$, not $b_{i0}$ or $b_{i1}$; this is ensured by the MRT design. The random noise $\epsilon_{it+1}$ is assumed to be independent of $(H_{it}, A_{it}, b_{0i}, b_{1i})$ and follows $N(0,\sigma_\epsilon^2)$. $f_0(H_{it})$ and $f_1(H_{it})$ can include possibly endogenous covariates $X_{it}$ and lagged outcomes such as $Y_{it}$.

Equation \eqref{eq:proposed-model} along with the above assumptions completely specifies the conditional distribution of the outcome $Y_{it+1}$ conditional on $b_{0i}, b_{1i}, H_{it}, A_{it}$. It implies the following treatment effect that is conditional on the random effects
\begin{align}
E(Y_{it+1} \mid b_{0i}, b_{1i}, H_{it}, A_{it}=1) - E(Y_{it+1} \mid b_{0i}, b_{1i}, H_{it}, A_{it}=0) = f_1(H_{it})^T \beta_1 + g_1(H_{it})^T b_{1i}. \label{eq:trt-eff-conditional}
\end{align}
Furthermore due to endogeneity, it is likely that
\begin{align}
E(Y_{it+1} \mid H_{it}, A_{it}=1) - E(Y_{it+1} \mid H_{it}, A_{it}=0) \neq f_1(H_{it})^T \beta_1. \label{eq:trt-eff-marginal}
\end{align}
In other words, the treatment effect \eqref{eq:trt-eff-conditional} implied by model \eqref{eq:proposed-model} is interpreted as \textit{conditional-on-the-random-effect}; $\beta = (\beta_0^T, \beta_1^T)^T$ does not have a \textit{marginal} interpretation.
A similar point for when there is no treatment has been extensively discussed in Section \ref{sec:explain-issue}.

The above model provides the distribution of $Y_{it+1}$ conditional on $(b_{0i}, b_{1i}, H_{it}, A_{it})$ as opposed to conditional on $(b_{0i}, b_{1i}, X_{it}, A_{it})$.  Thus  $\beta_1$ in \eqref{eq:trt-eff-conditional} has a causal interpretation even  when the randomization probability for  $A_{it}$  depends on $H_{it}$ in an MRT.   Likelihood-based inference and model fitting through standard LMM software can be conducted as described below. Note that since $f_0(H_{it})$ and $f_1(H_{it})$ can include lagged outcomes, the dependence between outcomes is explicitly modeled in \eqref{eq:proposed-model}. The purpose of introducing random effects here is mainly to model the between-person heterogeneity.

To estimate the conditional-on-the-random-effect $\beta$, we make an additional conditional independence assumption. The \textit{conditional independence assumption} is
\begin{equation}
X_{it} \perp (b_{0i}, b_{1i}) \mid H_{it-1}, A_{it-1}, Y_{it}. \label{eq:independence-assumption}
\end{equation}
This does allow $X_{it}$ to be endogenous, 
but the endogenous covariate $X_{it}$ can only depend on the random effects through the variables observed prior to $X_{it}$: $H_{it-1}, A_{it-1}$, and $Y_{it}$.
If the only endogenous covariates are functions of prior treatments and prior outcomes, then assumption \eqref{eq:independence-assumption} automatically holds.
In general, assumption \eqref{eq:independence-assumption} needs to be verified from the domain science perspective. We discuss this assumption in the context of HeartSteps in Section \ref{sec:data-analysis}.

Assumption \eqref{eq:independence-assumption} allows us to decompose the likelihood. This likelihood decomposition will provide a justification for the use of estimators from standard LMM software. 
Denote by $X_i$, $A_i$ and $Y_i$ the vectors of observations for individual $i$, and $X$, $A$ and $Y$ the collection of observations for all individuals. Denote by $b_i =(b_{0i}, b_{1i})$. Suppose $G$, the covariance matrix of the random effects, is parametrized by $\theta$. The joint likelihood of the observed data, $\cL(\alpha, \beta, \theta, \sigma_\epsilon \mid X, A, Y)$, can be written as
\begin{align}
\prod_i p(X_i, A_i, Y_i \mid \alpha, \beta, \theta, \sigma_\epsilon) &= \prod_i \int p(X_i, A_i, Y_i \mid b_i ; \alpha, \beta, \theta, \sigma_\epsilon) dF(b_i) \nonumber \\
&= \prod_i \Big\{ \int \prod_t
p(X_{it} \mid H_{it-1}, A_{it-1}, Y_{it}, b_i)
p(A_{it} \mid H_{it}, b_i) \nonumber \\
& \qquad \qquad \times p(Y_{it+1} \mid H_{it}, A_{it}, b_i ; \alpha, \beta, \theta, \sigma_\epsilon) dF(b_i) \Big\}. \label{eq:lkd-1}
\end{align}
By the conditional independence assumption \eqref{eq:independence-assumption} and given that $A_{it}$ is randomized conditional on $H_{it}$, the joint likelihood in \eqref{eq:lkd-1} becomes
\begin{align}
\cL(\alpha, \beta, \theta, \sigma_\epsilon \mid X, A, Y) & = \Big\{ \prod_i \prod_t p(X_{it} \mid H_{it-1}, A_{it-1}, Y_{it})
p(A_{it} \mid H_{it}) \Big\} \cL_1(\alpha, \beta, \theta, \sigma_\epsilon \mid X, A, Y), 
\label{eq:lkd-2}
\end{align}
where
\begin{align}
\cL_1(\alpha, \beta, \theta, \sigma_\epsilon \mid X, A, Y) = \prod_i \Big\{ \int \prod_t p(Y_{it+1} \mid H_{it}, A_{it}, b_i ; \alpha, \beta, \theta, \sigma_\epsilon) dF(b_i) \Big\}. \label{eq:partial-lkd}
\end{align}

Because the first factor on the right hand side of \eqref{eq:lkd-2} does not involve $(\alpha, \beta, \theta, \sigma_\epsilon)$, any inference for $(\alpha, \beta, \theta, \sigma_\epsilon)$ that is based on the joint likelihood $\cL(\alpha, \beta, \theta, \sigma_\epsilon \mid X, A, Y)$ can be equivalently based on the partial likelihood $\cL_1(\alpha, \beta, \theta, \sigma_\epsilon \mid X, A, Y)$. Observe that $\cL_1(\alpha, \beta, \theta, \sigma_\epsilon \mid X, A, Y)$ is actually the likelihood function for a standard LMM where $X_{it}$ and $A_{it}$ are treated as fixed covariates.
Thus, the maximum likelihood estimators that are obtained through standard LMM software are valid maximum likelihood estimators for the joint likelihood $\cL(\alpha, \beta, \theta, \sigma_\epsilon \mid X, A, Y)$ under the conditional independence assumption, and \eqref{eq:lmm-ee} with $X$ redefined to include the treatment indicator is a likelihood score equation for $\beta$ in the conditional-on-the-random-effect model. Note that even though the form of \eqref{eq:lmm-ee} appears to indicate estimation of a regression coefficient in a marginal model, this is a false impression in the case of endogenous covariates. 
Furthermore, recall that restricted maximum likelihood (REML)
estimation can be viewed as maximum \textit{a posteriori} in a Bayesian hierarchical model \citep{laird1982random}. This latter interpretation continues to hold for the REML estimators obtained through standard LMM software when there are endogenous covariates. In addition, it can be shown that the empirical Bayes predictor of the random effects $\hat{b}_i$ obtained through standard LMM software is valid empirical Bayes predictor for model \eqref{eq:proposed-model} with endogenous covariates. We include proofs of these claims in the Appendix.

The conditional independence assumption \eqref{eq:independence-assumption} is similar to an assumption used by 
\citet{sitlani2012}. \citet{sitlani2012} aimed to use an LMM to assess causal effects in the context of noncompliance in surgical trials. They assumed conditional independence between the treatment assignment and the random effect given the observed history. This assumption allowed them to decompose the likelihood as is done above and thus use standard LMM estimators.

It is worth noting, as pointed out by a reviewer, that if the analyst poses a model as \eqref{eq:proposed-model} but without the $A_{it} g_1(H_{it})^T b_{1i}$ term (i.e., the random effect in the model does not interact with $A_{it}$), then \eqref{eq:trt-eff-marginal} becomes an equality. In other words, in this case $\beta_1$ recovers its marginal interpretation
\begin{align*}
E(Y_{it+1} \mid H_{it}, A_{it}=1) - E(Y_{it+1} \mid H_{it}, A_{it}=0) = f_1(H_{it})^T \beta_1,
\end{align*}
and furthermore it can be interpreted marginally over $H_{it} \setminus f_1(H_{it})$:
\begin{align}
E \big\{ E(Y_{it+1} \mid H_{it}, A_{it}=1) - E(Y_{it+1} \mid H_{it}, A_{it}=0) \mid f_1(H_{it}) \big\} = f_1(H_{it})^T \beta_1. \label{eq:beta-causal-excursion-interpretation}
\end{align}
Note that $\beta_0$ still has only the conditional-on-the-random-effect interpretation. In absence of $b_{1i}$, the conditional independence assumption \eqref{eq:independence-assumption} becomes
\begin{align*}
  X_{it} \perp b_{0i} \mid H_{it-1}, A_{it-1}, Y_{it};
\end{align*}
this assumption justifies the  use of  over-the-counter LMM software's via the likelihood factorization \eqref{eq:lkd-2}.




\section{Simulation}
\label{sec:simulation}

In the simulation, we considered three generative models (GMs), in all of which the covariate is endogenous. 
In the first two GMs, the endogenous covariate $X_{it}$ equals the previous outcome $Y_{it}$ plus some random noise, so the conditional independence assumption \eqref{eq:independence-assumption} is valid. In GM 3, the endogenous covariate depends directly on $b_i$, so the assumption \eqref{eq:independence-assumption} is violated. Details of the generative models are described in the following. 

In GM 1, we considered a simple case with only a random intercept and a random slope for $A_{it}$, so that $Z_{it}^{(0)} = Z_{it}^{(2)} = 1$ in model \eqref{eq:proposed-model}. The outcome is generated as $Y_{it+1} = \alpha_0 + \alpha_1 X_{it} + b_{i0} + A_{it}(\beta_0 + \beta_1 X_{it} + b_{i2}) + \epsilon_{it+1}$. The random effects $b_{i0} \sim N(0, \sigma_{b0}^2)$ and $b_{i2} \sim N(0, \sigma_{b2}^2)$ are independent of each other. We generated the covariate to be $X_{i1} \sim N(0,1)$, $X_{it} = Y_{it} + N(0,1)$ for $t \geq 2$. The randomization probability $p_t$ is constant 1/2. The exogenous noise $\epsilon_{it+1} \sim N(0,\sigma_\epsilon^2)$.

In GM 2, we considered the case where $Z_{it}^{(0)} = Z_{it}^{(2)} = (1, X_{it})$, and the randomization probability is time-varying. The outcome is generated as $Y_{it+1} = \alpha_0 + \alpha_1 X_{it} + b_{i0} + b_{i1} X_{it} + A_{it}(\beta_0 + \beta_1 X_{it} + b_{i2} + b_{i3} X_{it}) + \epsilon_{it+1}$. The random effects $b_{ij} \sim N(0, \sigma_{bj}^2)$, $0\leq j \leq 3$, are independent of each other. We generated the covariate to be $X_{i1} \sim N(0,1)$, $X_{it} = Y_{it} + N(0,1)$ for $t \geq 2$. The randomization probability depends on $X_{it}$: $p_{t}=0.7 \cdot \ind(X_{it}>-1.27)+0.3 \cdot \ind(X_{it}\leq -1.27)$. Here $\ind(\cdot)$ represents the indicator function, and the cutoff $-1.27$ was chosen so that $p_t$ equals 0.7 or 0.3 each for about half of the time. The exogenous noise $\epsilon_{it+1} \sim N(0,\sigma_\epsilon^2)$.

GM 3 is the same as GM 1, except that the covariate $X_{it}$ depends directly on $b_i$: $X_{i1} \sim N(b_{i0},1)$, $X_{it} = Y_{it} + N(b_{i0},1)$ for $t \geq 2$.

We chose the parameter values as follows: $\alpha_0 = -2$, $\alpha_1 = -0.3$, $\beta_0 = 1$, $\beta_1 = 0.3$, $\sigma_{b0}^2 = 4$, $\sigma_{b1}^2 = 1/4$, $\sigma_{b2}^2 = 1$, $\sigma_{b3}^2 = 1/4$, $\sigma_\epsilon^2 = 1$.

For each of the three GMs, we simulated for sample size $n=30,100,200$ and the number of observations per individual $T_i = T = 10, 30$. Each setting was replicated 1,000 times. The estimation was done using the R package \textsf{lmer} \citep{lmer} for standard LMM, and 95\% confidence interval was computed based on the $t$ distribution with degrees of freedom obtained by Satterthwaite approximation \citep{satterthwaite1941synthesis}, which is implemented in the R package \textsf{lmerTest} \citep{lmerTest}.
Bias, standard deviation (sd) and coverage probability (cp) of 95\% nominal confidence interval for the estimated $\beta_0$ and $\beta_1$ are presented in Table \ref{tab:simulation-result}.
As expected, the estimators are consistent for GM 1 and GM 2, and they are inconsistent for GM 3 because of the violation of the conditional independence assumption \eqref{eq:independence-assumption}. For GM 1 and GM 2, the confidence interval coverage probability can be slightly lower than the nominal level for some of the parameters for small $n$ or small $T$, but it gets back to the nominal level as the sample size or total number of time points gets larger. Additional simulation results for more choices of $n$ and $T$, the performance of estimated $\alpha_0$, $\alpha_1$, and variance components $\sigma_{bj}^2, 0\leq j\leq 3$ and $\sigma_\epsilon^2$ are in the Appendix, and the conclusion is similar to the results for the $\beta$'s as shown here.

\begin{table}[htbp]
\centering

\begin{tabular}{rrrrrrrrr}
  \hline
& & & \multicolumn{3}{c}{$\beta_0$} & \multicolumn{3}{c}{$\beta_1$} \\
\cmidrule(lr){4-6} \cmidrule(lr){7-9} 
GM & $T$ & $n$ & bias & sd & cp & bias & sd & cp \\ 
  \hline
    & & 30 & -0.001 & 0.249 & 0.943 & 0.002 & 0.091 & 0.897 \\ 
   & & 100 & -0.003 & 0.135 & 0.941 & -0.001 & 0.049 & 0.898 \\ 
  \multirow{-3}{*}{\centering 1} & \multirow{-3}{*}{\centering 10} & 200 & -0.001 & 0.096 & 0.926 & -0.001 & 0.034 & 0.899 \\
   \hline
    &  & 30 & -0.002 & 0.206 & 0.946 & 0.001 & 0.053 & 0.913  \\
   &  & 100 & -0.005 & 0.112 & 0.949 & -0.001 & 0.028 & 0.935 \\
  \multirow{-3}{*}{\centering 1} & \multirow{-3}{*}{\centering 30} & 200 & 0.000 & 0.081 & 0.944 & -0.001 & 0.022 & 0.902  \\
   \hline
    &  & 30 & -0.010 & 0.269 & 0.939 & -0.004 & 0.105 & 0.903 \\
   &  & 100 & 0.009 & 0.145 & 0.933 & -0.001 & 0.056 & 0.915  \\
  \multirow{-3}{*}{\centering 2} & \multirow{-3}{*}{\centering 10} & 200 & -0.008 & 0.105 & 0.931 & -0.002 & 0.038 & 0.934 \\
   \hline
    &  & 30 & -0.006 & 0.216 & 0.943 & -0.001 & 0.070 & 0.939  \\
   &  & 100 & 0.006 & 0.115 & 0.947 & -0.001 & 0.039 & 0.948  \\
  \multirow{-3}{*}{\centering 2} & \multirow{-3}{*}{\centering 30} & 200 & -0.004 & 0.084 & 0.935 & -0.000 & 0.027 & 0.940  \\
   \hline
    &  & 30 & -0.048 & 0.245 & 0.949 & -0.043 & 0.075 & 0.725  \\
   &  & 100 & -0.060 & 0.134 & 0.927 & -0.047 & 0.041 & 0.548  \\
  \multirow{-3}{*}{\centering 3} & \multirow{-3}{*}{\centering 10} & 200 & -0.052 & 0.095 & 0.907 & -0.046 & 0.029 & 0.355  \\
   \hline
    &  & 30 & -0.023 & 0.207 & 0.946 & -0.017 & 0.041 & 0.847  \\
   &  & 100 & -0.028 & 0.112 & 0.942 & -0.019 & 0.022 & 0.762  \\
  \multirow{-3}{*}{\centering 3} & \multirow{-3}{*}{\centering 30} & 200 & -0.024 & 0.079 & 0.941 & -0.019 & 0.015 & 0.628  \\
   \hline
\end{tabular}

\caption{Bias, standard deviation (sd) and coverage probability (cp) of 95\% nominal confidence interval for estimated $\beta_0$ and $\beta_1$ in the simulation study. $n$ denotes sample size; $T$ denotes total number of observations for each individual; GM denotes generative model. The result is based on 1,000 replicates for each setting.}
\label{tab:simulation-result}
\end{table}

\section{Illustrative data analysis of HeartSteps}
\label{sec:data-analysis}

\subsection{Data and model assumptions}

As described in Section \ref{subsec:motivating-example}, HeartSteps \citep{klasnja2018} is a 6-week micro-randomized trial of an mHealth intervention to encourage activity among sedentary adults. The following analysis focuses on the time-varying treatment consisting of contextually-tailored activity suggestions.

Prior to the randomization at each time point, software on the smartphone determined whether an individual is \textit{available} for treatment at the time. If the activity recognition on the phone determined that an individual was operating a vehicle, the individual was considered unavailable for safety reasons. If an individual had just finished an activity bout in the prior 90 seconds, they were considered unavailable for treatment in order to minimize user burden and aggravation. Lastly, because the software on the server and smartphone required an internet connection to send a suggestion, if the smartphone did not have wireless connectivity the individual was deemed unavailable. At each of the five points each day for each individual, availability was assessed, the context was recorded, and if the individual was available then HeartSteps randomized to deliver an activity suggestion to the individual with probability 3/5. The sample for this analysis consisted of 7,540 time points from 37 individuals. The individuals were available for 6,061 (80.4\%) time points, unavailable due to no internet connection for 602 (8.0\%) time points, unavailable due to being detected as in transit for 841 (11.1\%) time points, and unavailable due to being detected to have just finished an activity bout in the prior 90 seconds for 36 (0.5\%) time points.

Let $A_{it}=1$ if an activity suggestion is delivered at time $t$ for individual $i$ and equal to $0$ otherwise. The proximal outcome $Y_{it+1}$ is the (log-transformed) 30-minute step count following time point $t$. We used three covariates in the model:
\begin{itemize}
\item $X_{it,1}$: day in the study for the time point $t$, coded as $0,1,\ldots,41$.
\item $X_{it,2}$: whether the individual was at home or work at time point $t$; $X_{it,2} = 1$ if at home or work, 0 if at some other location.
\item $X_{it,3}$: (log-transformed) 30-minute step count preceding time point $t$.
\end{itemize}
We specify model \eqref{eq:proposed-model} in the HeartSteps context as follows: $f_0(H_{it}) = (X_{it,1}, X_{it,2}, X_{it,3})$; $f_1(H_{it}) = (X_{it,1},$ $X_{it,2})$; the model contains a random intercept, $g_0(H_{it}) = 1$, and a random slope for $A_{it}$, $g_1(H_{it}) = 1$. We denote the availability status of individual $i$ at time $t$ by $I_{it}$ ($I_{it} = 1$ if available; $0$ otherwise). In the model, we multiply $A_{it}$ with $I_{it}$ to operationalize the notion that the treatment may only be delivered when the individual is available. Because the relationship between $Y_{it+1}$ and the $f_0(H_{it})$ can depend on the availability status, we included an interaction between $I_{it}$ and $f_0(H_{it})$. Thus, the LMM is given by
\begin{align}
Y_{it+1} & = \alpha_0 + \alpha_1 X_{it,1} + \alpha_2 X_{it,2} + \alpha_3 X_{it,3} + I_{it} (\tilde{\alpha}_0 + \tilde{\alpha}_1 X_{it,1} + \tilde{\alpha}_2 X_{it,2} + \tilde{\alpha}_3 X_{it,3}) + b_{0i} \nonumber \\
& \quad + A_{it} I_{it} (\beta_0 + \beta_1 X_{it,1} + \beta_2 X_{it,2} + b_{1i}) + \epsilon_{it+1} \label{eq:model-data-analysis}
\end{align}
where $\epsilon_{it+1} \sim N(0,\sigma_\epsilon^2)$, and the random effects $(b_{0i}, b_{1i}) \sim N(0, G)$ with $G$ being a $2\times 2$ variance-covariance matrix. $b_{0i}$ accounts for the between-individual variation in the 30-minute step count under no treatment, and $b_{1i}$ accounts for the between-individual variation in the treatment effect on the 30-minute step count.

In model \eqref{eq:model-data-analysis}, $X_{it,2}$, $X_{it,3}$ and $I_{it}$ are possibly endogenous.
Location, $X_{it,2}$, is most likely exogenous but might be endogenous because the number of steps an individual took following a prior time point, combined with the location s/he was at then, might be predictive of whether s/he would be at home/work or other places at the subsequent time point. Prior time $t$ 30-minute step count,
$X_{it,3}$, might be correlated with 30-minute step count after time $t-1$, $Y_{it}$, because an individual might walk less if s/he had already walked earlier in the day.
For the availability status $I_{it}$, unavailability due to being in transit is likely exogenous but may be endogenous for a reason similar to that of location, $X_{it,2}$. Unavailability due to having just finished an activity bout may be endogenous for a reason similar to that of prior time $t$ 30-minute step count, $X_{it,3}$.
We argue that the conditional independence assumption \eqref{eq:independence-assumption} is plausible for all three variables. 
For location, $X_{it,2}$, because the enrollment criterion required each individual to either have a full-time daytime job or be a student, the time-varying location of such individuals with regular schedule is unlikely to depend on some unmeasured baseline factors (i.e., the random effects) that impact step count.
For prior time $t$ 30-minute step count, $X_{it,3}$, the impact of random effects should be largely explainable through earlier outcomes and covariates, as those are also step counts but just for other time windows. 
For $I_{it}$, most of the unavailability (1443/1479) instances are due to being in transit or loss of internet connection; the conditional independence is likely to approximately hold for $I_{it}$ for a similar reason to that of $X_{it,2}$.

\subsection{Results}

We fitted model \eqref{eq:model-data-analysis} using the R package \textsf{lmer} \citep{lmer} for standard LMM, because standard LMM yields valid estimators under the conditional independence assumption \eqref{eq:independence-assumption}. 

The first three columns in Table \ref{tab:data-analysis-result} show the estimated fixed effects with 95\% confidence interval and the estimated variance components. The estimated variance for $b_{1i}$ is extremely small and the estimated correlation between $b_{0i}$ and $b_{1i}$ is 1.000, suggesting that we might not have enough data to fit two separate random effects so the fitting collapsed onto a linear combination of the two. We conducted the likelihood ratio test for nonzero variance of $b_{1i}$, and the p-value was 0.72. Note that likelihood ratio tests for nonzero variance components can be conservative because the null value ($\var(b_{1i}) = 0$) is on the boundary of the parameter space \citep{self1987asymptotic, stram1994variance, crainiceanu2004likelihood}, and we are just using this test and the critical value as a guideline. The result suggests that the potential heterogeneity in the treatment effect may not be large enough to be detected from the data. Model fit of \eqref{eq:model-data-analysis} with $b_{1i}$ removed is presented in the last two columns in Table \ref{tab:data-analysis-result}.

The estimated treatment effects, which are conditional on the observed history and the unobserved random effects, are similar from both model fits in the point estimates as well as the confidence intervals.
The data indicates that, for an individual, the treatment has a positive effect at the beginning of the study ($\hat\beta_0 > 0$), and the effect decreased over time ($\hat\beta_1 < 0$). This is likely due to the individual's habituation to the activity suggestions, which is consistent with the exit interviews reported by \citet{klasnja2018} in which individuals reported that ``the suggestions became boring after 2--4 weeks''. On the other hand, the data indicates no moderating influence of location (whether an individual was at home/work or some other place) on the treatment effect for an individual.


\begin{table}[htbp]
\centering
\begin{tabular}{rrcrc}
  \hline
  & \multicolumn{2}{c}{Model with $b_{1i}$} & \multicolumn{2}{c}{Model without $b_{1i}$} \\
  \cmidrule(lr){2-3} \cmidrule(lr){4-5}
coefficient & estimate & 95\% CI & estimate & 95\% CI\\ 
  \hline
$\alpha_0$ & 1.990 & (\phm1.643, \phm2.338) & 1.997 & (\phm1.646, \phm2.348) \\ 
  $\alpha_1$ & -0.009 & (-0.021, \phm0.002) & -0.009 & (-0.021, \phm0.002) \\ 
  $\alpha_2$ & 0.851 & (\phm0.238, \phm1.465) & 0.840 & (\phm0.226, \phm1.453) \\ 
  $\alpha_3$ & 0.539 & (\phm0.495, \phm0.583) & 0.537 & (\phm0.493, \phm0.582) \\ 
  $\tilde{\alpha}_0$ & -0.177 & (-0.586, \phm0.232) & -0.182 & (-0.591, \phm0.228) \\ 
  $\tilde{\alpha}_1$ & 0.008 & (-0.006, \phm0.023) & 0.008 & (-0.007, \phm0.023) \\ 
  $\tilde{\alpha}_2$ & -0.871 & (-1.522, -0.221) & -0.863 & (-1.514, -0.212) \\ 
  $\tilde{\alpha}_3$ & -0.156 & (-0.206, -0.107) & -0.154 & (-0.204, -0.104) \\ 
  $\beta_0$ & 0.415 & (\phm0.105, \phm0.724) & 0.410 & (\phm0.100, \phm0.719) \\ 
  $\beta_1$ & -0.017 & (-0.028, -0.005) & -0.017 & (-0.028, -0.005) \\ 
  $\beta_2$ & 0.122 & (-0.156, \phm0.400) & 0.130 & (-0.148, \phm0.408) \\ 
   \hline
   \hline
   $\var(b_{0i})$          & 0.160   & & 0.182 &       \\
   $\var(b_{1i})$          & 0.003    & & - &     \\
 $\corr(b_{0i}, b_{1i})$ & 1.000    & & - &     \\
$\var(\epsilon_{it+1})$                  & 7.138  & & 7.139 &        \\
   \hline
\end{tabular}
\caption{Estimated coefficients and 95\% confidence interval for model \eqref{eq:model-data-analysis} of HeartSteps data. Estimators are obtained using R package \textsf{lmer}, and the 95\% confidence interval are based on $t$ distribution with Satterthwaite approximation implemented in R package \textsf{lmerTest}.}
\label{tab:data-analysis-result}
\end{table}

As a point of contrast, we also analyzed the data using the weighted and centered least-squares (WCLS) estimator in \citet{boruvka2017} for a related but different model. We used WCLS to estimate $\psi = (\psi_0, \psi_1, \psi_2)$ in the following model:
\begin{align}
  E \big\{ E(Y_{it+1} \mid H_{it}, A_{it} = 1) - E(Y_{it+1} \mid H_{it}, A_{it} = 0) \mid X_{it,1}, X_{it,2}, I_{it} = 1 \big\} = \psi_0 + \psi_1 X_{it,1} + \psi_2 X_{it,2}. \label{eq:model-data-analysis-wcls}
\end{align}
\citet{boruvka2017} called \eqref{eq:model-data-analysis-wcls} the causal excursion effect; $\psi$ is marginal over both the random effects and $H_{it} \setminus \{X_{it,1}, X_{it,2}\}$, which is different from $\beta$ in \eqref{eq:model-data-analysis}. We used $\gamma_0 + \gamma_1 X_{it,1} + \gamma_2 X_{it,2} + \gamma_3 X_{it,3}$ as the working model for $E(Y_{it+1} \mid H_{it}, A_{it} = 0, I_{it} = 0)$ in WCLS; this working model does not need to be correctly specified to guarantee the consistent of the estimator for $\psi$. The estimated $\psi$ and the 95\% confidence interval are listed in Table \ref{tab:data-analysis-result-wcls}. Although $\beta$ and $\psi$ are different estimands with different interpretation, their estimated value and confidence interval are qualitatively similar. These results are consistent with the comments made in the last paragraph regarding the direction of how different variables moderate the treatment effect.

\begin{table}[htbp]
\centering
\begin{tabular}{rrl}
  \hline
 coefficient & estimate & 95\% CI \\ 
  \hline
  $\psi_0$ & 0.454 & (\phm0.156, \phm0.753) \\ 
  $\psi_1$ & -0.018 & (-0.029, -0.006) \\ 
  $\psi_2$ & 0.096 & (-0.219, \phm0.410) \\ 
   \hline
\end{tabular}
\caption{Estimated coefficients and 95\% confidence interval for model \eqref{eq:model-data-analysis-wcls} using WCLS estimator in \citet{boruvka2017}.}
\label{tab:data-analysis-result-wcls}
\end{table}

\section{Discussion}
\label{sec:discussion}

Linear mixed models (LMM) were originally developed for settings with fixed covariates, and it has been natural for researchers to think about the induced marginal model when building and interpreting the fixed effects in LMM. In this paper, we review related literature on the potential bias that would arise when including endogenous covariates into LMM. We argued that the fundamental issue in LMM with endogenous covariates is that the fixed effects, including the treatment effect, will only have a conditional-on-the-random-effect interpretation, and the marginal interpretation is no longer valid. In terms of estimation for LMM with endogenous covariates, we introduced a conditional independence assumption, and showed that under this assumption standard LMM software can still be used to obtain valid estimator of the fixed effects and the variance components, as well as valid prediction of the random effects. We used an LMM to model the  effect of sequentially assigned treatment in HeartSteps MRT in which the covariates are likely endogenous, and we discussed the plausibility of the conditional independence assumption for these covariates.

The potential bias resulting from endogenous covariates in the without-treatment longitudinal setting has been known for decades since \citet{pepeanderson1994}. However, it was quite surprising to us that in the MRT setting, this issue occurs even with randomized treatment with constant randomization probability (no confounding). The method in this paper utilizes the randomization to the extent that the treatment indicator $A_{it}$ automatically satisfies a conditional independence assumption similar to \eqref{eq:independence-assumption}. Furthermore, \eqref{eq:proposed-model} is a mechanistic model for the outcome, which implies that how well the estimated $\beta$ approximates the true treatment effect is contingent on how well the mechanistic model approximates the true data generating distribution. When the marginal treatment effect is of interest, there are many tools in causal inference that consistently estimate the effect with a possibly misspecified nuisance model \citep{robins1994snmm, robins2000marginal, hernan2001marginal, brumback2003intensity, goetgeluk2008conditional, boruvka2017}. It is an open question whether the randomization can be further leveraged in LMM to increase robustness to misspecified nuisance models.

The inclusion of endogenous covariates to an LMM implies that the fixed effects should only be interpreted  as  conditional on an individual. Thus, a future research question is to develop estimation methods for the parameters in the marginal mean model that are coherent with fixed effect parameters in an LMM where there are endogenous covariates. Related work in generalized linear mixed models but with exogenous covariates includes \citet{heagerty1999marginally}, \citet{heagerty2000marginalized}, and \citet{larsen2000interpreting}.

In a standard LMM with exogenous covariates, the empirical best linear unbiased predictor (eBLUP) equals the empirical Bayes estimator where a noninformative prior is imposed on the fixed effect and the variance components are estimated through REML \citep{lindley1972bayes, dempfle1977comparison}. In Section \ref{sec:model} we showed through partial likelihood argument that the empirical Bayes estimator of random effects from standard LMM is still a valid empirical Bayes estimator in the case of endogenous covariates. However, it is unknown whether it
is still eBLUP  absent further assumptions. 

Along the same lines, in a standard LMM the restricted maximum likelihood (REML) estimator of the variance components can be viewed as the maximum \textit{a posteriori} estimator in a Bayesian hierarchical model \citep{laird1982random}, and in Section \ref{sec:model} we showed that this latter interpretation is valid for the REML estimators obtained through standard LMM software when there are endogenous covariates. Another interpretation of the REML estimator in a standard LMM is the maximizer for the likelihood of linear combinations of the outcome that is orthogonal to the fixed effects. It is unknown whether this interpretation continues to hold for the endogenous covariate case.


In the literature, there has been work on handling endogenous covariates in longitudinal data via jointly modeling of the covariate process and the outcome process, which could be alternative approaches to the method proposed in this paper for situations where the conditional independence assumption is questionable. Note that each of these alternative approaches require certain assumptions on the covariate process, and these assumptions themselves need to be verified in the context of each application. For example, \citet{miglioretti2004marginal} modeled the covariate process, and assumed that $X_{it} \perp b_i \mid X_{i1}, X_{i2}, \ldots, X_{it-1}$. \citet{roy2006conditional} proposed to model the distribution of covariates given the history to infer the dependence of a Poisson process outcome on the endogenous covariates. \citet{sitlani2012} proposed to use joint modeling for analyzing the effect of a surgical trial (where the time-varying treatment is a jump process) under noncompliance. \citet{shardell2018joint} proposed to use a joint model approach, by assuming either that the distribution of $X_{it}$ can be correctly modeled, or that the endogenous covariate is the lagged outcome.


\appendix

\section{Estimation and prediction through standard LMM software} 

In this Appendix, we provide a proof for the claims in Section \ref{sec:model} that maximum likelihood estimators, maximum \textit{a posterior} estimators, and the empirical Bayes prediction of the random effects can be obtained through standard LMM software.

\subsection{Estimation of fixed effects and variance components}
\label{subsec:estimation}

This subsection focuses on estimation of the fixed effects $\alpha$ and $\beta$ and the variance components $\theta$ and $\sigma_\epsilon^2$ in model \eqref{eq:proposed-model}. 

That the maximum likelihood estimator for the fixed effects and the variance component can be obtained through standard LMM software is immediate from the likelihood factorization \eqref{eq:lkd-2}.


The restricted maximum likelihood (REML) estimator of the variance components $\theta$ and $\sigma_\epsilon$ in a standard LMM can be obtained through Bayesian maximum \textit{a posteriori} (MAP) estimation with a non-informative prior on the fixed effects $\alpha,\beta$ \citep{laird1982random, searle1992variance}. For our case, the marginal likelihood for $\theta, \sigma_\epsilon$, where $\alpha$ and $\beta$ are integrated over with respect to non-informative priors $p(\alpha)$ and $p(\beta)$, is
\begin{align}
L(\theta, \sigma_\epsilon \mid X_i, A_i, Y_i, 1\leq i \leq n) = \int p(\alpha) p(\beta) \prod_i p(X_i, A_i, Y_i \mid \alpha, \beta, \theta, \sigma_\epsilon) d\alpha d\beta, \nonumber
\end{align}
which by \eqref{eq:lkd-2} equals
\begin{align}
& \prod_i \Big\{ \prod_t p(X_{it} \mid H_{it-1}, A_{it-1}, Y_{it})
p(A_{it} \mid H_{it}) \Big\} \nonumber \\
& \qquad \times \int p(\alpha) p(\beta) \prod_i \Big\{ \int \prod_t p(Y_{it+1} \mid H_{it}, A_{it}, b_i ; \alpha, \beta, \theta, \sigma_\epsilon) dF(b_i) \Big\} d\alpha d\beta \nonumber \\
& \propto \int p(\alpha) p(\beta) \prod_i \Big\{ \int \prod_t p(Y_{it+1} \mid H_{it}, A_{it}, b_i ; \alpha, \beta, \theta, \sigma_\epsilon) dF(b_i) \Big\} d\alpha d\beta.
\label{eq:lkd-3}
\end{align}
Expression \eqref{eq:lkd-3} is the marginal likelihood for $\theta, \sigma_\epsilon$ in a standard LMM; hence, the MAP estimator of the variance components can be obtained through standard LMM fitting procedure with the REML option.


\subsection{Prediction of random effects}
\label{subsec:prediction}

Prediction of random effects in a standard LMM is through best linear unbiased predictors (BLUPs, \citet{henderson1975blup}), which can be alternatively derived as empirical Bayes estimates using REML estimator of the variance components and fixed effects \citep{lindley1972bayes, dempfle1977comparison}.

Denote by $b = (b_1, \ldots, b_n)$, $X = (X_1, \ldots, X_n)$, $A = (A_1, \ldots, A_n)$, and $Y = (Y_1, \ldots, Y_n)$. In our proposed model, the posterior distribution of $b$ is
\begin{align}
p(b \mid X, A, Y; \theta, \sigma_\epsilon) = \frac{p(b, X, A, Y \mid \theta, \sigma_\epsilon)}{p(X, A, Y \mid \theta, \sigma_\epsilon)}. \label{eq:posterior-1}
\end{align}
We omit the notational dependence on $\theta,\sigma_\epsilon$ hereafter. Let $p(\alpha)$ and $p(\beta)$ denote the prior distribution of $\alpha$ and $\beta$. The numerator of the right hand side of \eqref{eq:posterior-1} equals
\begin{align}
\int p(b,X,A,Y,\alpha,\beta)d\alpha d\beta & = \int p(\alpha)p(\beta) \prod_i p(b_i) \prod_t p(X_{it} \mid H_{it-1}, A_{it-1}, Y_{it}, b_i, \alpha, \beta) \nonumber \\
& \quad \times p(A_{it} \mid H_{it}, b_i, \alpha, \beta) p(Y_{it+1} \mid H_{it}, A_{it}, b_i ; \alpha, \beta)
d\alpha d\beta \nonumber \\
& = \Big\{ \prod_i \prod_t p(X_{it} \mid H_{it-1}, A_{it-1}, Y_{it}) p(A_{it} \mid H_{it}) \Big\} \nonumber \\
& \quad\times \int p(\alpha)p(\beta) \prod_i p(b_i) \prod_t p(Y_{it+1} \mid H_{it}, A_{it}, b_i ; \alpha, \beta)
d\alpha d\beta,
\end{align}
where the last equality follows from the conditional independence assumption and the randomization of $A_{it}$. The denominator of the right hand side of \eqref{eq:posterior-1} is $\int \int p(b,X,A,Y,\alpha,\beta)d\alpha d\beta db$. Thus, the posterior distribution \eqref{eq:posterior-1} equals
\begin{align}
\frac{ \int p(\alpha)p(\beta) \prod_i p(b_i) \prod_t p(Y_{it+1} \mid H_{it}, A_{it}, b_i ; \alpha, \beta)
d\alpha d\beta }{ \int p(\alpha)p(\beta) \prod_i p(b_i) \prod_t p(Y_{it+1} \mid H_{it}, A_{it}, b_i ; \alpha, \beta)
d\alpha d\beta db },
\end{align}
which is the posterior distribution of $b$ in a standard LMM when $X$ and $A$ are treated as fixed or exogenous.

Therefore, the Bayesian MAP estimator of $b$ can be obtained through standard LMM fitting procedure. Along the same line, the empirical Bayes estimator of $b$ with plug-in variance component estimates can also be obtained through standard LMM.

\section{Additional simulation results}

In the additional simulation results, we included simulations for sample size $n=30,50,100,200$ and the number of observations per individual $T_i = T = 10,20,30$. Each setting was replicated 1,000 times. Bias, standard deviation (sd) and coverage probability (cp) of 95\% nominal confidence interval for the estimated fixed effects ($\beta$'s and $\alpha$'s) are presented in Table \ref{tab:simulation-result-full}. Table \ref{tab:simulation-result-varcomp} presents the bias and standard deviation for the estimated variance components $\sigma_{bj}^2,0 \leq j \leq 3$ and $\sigma_\epsilon^2$. For GM 1 and GM 3, the model doesn't include $b_{i1}$ and $b_{i3}$, so the variance components only include $\sigma_{b0}^2, \sigma_{b2}^2$ and $\sigma_\epsilon^2$. Conclusion to Section \ref{sec:simulation} can be made: for GM 1 and GM 2, the variance components are consistently estimated, whereas for GM 3 the estimators are inconsistent. Again, this is due to violation of the conditional independence assumption \eqref{eq:independence-assumption} in GM 3.

\afterpage{
\begin{landscape}

\begin{table}[htbp]
\centering

\begin{tabular}{rrrrrrrrrrrrrrr}
  \hline
& & & \multicolumn{3}{c}{$\beta_0$} & \multicolumn{3}{c}{$\beta_1$} & \multicolumn{3}{c}{$\alpha_0$} & \multicolumn{3}{c}{$\alpha_1$} \\
\cmidrule(lr){4-6} \cmidrule(lr){7-9} \cmidrule(lr){10-12} \cmidrule(lr){13-15}
GM & $T$ & $n$ & bias & sd & cp & bias & sd & cp & bias & sd & cp & bias & sd & cp \\ 
  \hline
 & & 30 & -0.001 & 0.249 & 0.943 & 0.002 & 0.091 & 0.897 & -0.021 & 0.377 & 0.951 & -0.002 & 0.065 & 0.915 \\ 
 & & 50 & -0.002 & 0.187 & 0.953 & -0.001 & 0.068 & 0.897 & -0.019 & 0.295 & 0.947 & -0.001 & 0.048 & 0.930 \\ 
& & 100 & -0.003 & 0.135 & 0.941 & -0.001 & 0.049 & 0.898 & -0.011 & 0.210 & 0.949 & -0.001 & 0.033 & 0.920 \\ 
  \multirow{-3}{*}{\centering 1} & \multirow{-3}{*}{\centering 10} & 200  & -0.001 & 0.096 & 0.926 & -0.001 & 0.034 & 0.899 & -0.009 & 0.150 & 0.941 & 0.000 & 0.025 & 0.909 \\ 
   \hline
 & & 30 & -0.001 & 0.217 & 0.943 & 0.001 & 0.063 & 0.919 & -0.020 & 0.372 & 0.950 & -0.002 & 0.046 & 0.928 \\ 
 & & 50 & 0.001 & 0.168 & 0.947 & -0.000 & 0.048 & 0.916 & -0.018 & 0.288 & 0.945 & -0.002 & 0.034 & 0.935 \\ 
& & 100 & -0.002 & 0.117 & 0.950 & -0.000 & 0.035 & 0.906 & -0.010 & 0.207 & 0.946 & -0.000 & 0.025 & 0.930 \\ 
  \multirow{-3}{*}{\centering 1} & \multirow{-3}{*}{\centering 20} & 200  & -0.001 & 0.085 & 0.943 & -0.001 & 0.026 & 0.892 & -0.008 & 0.147 & 0.944 & 0.000 & 0.018 & 0.921 \\ 
   \hline
 & & 30 & -0.002 & 0.206 & 0.946 & 0.001 & 0.053 & 0.913 & -0.020 & 0.367 & 0.952 & -0.001 & 0.038 & 0.924 \\ 
 & & 50 & -0.000 & 0.160 & 0.949 & 0.001 & 0.040 & 0.930 & -0.017 & 0.288 & 0.945 & -0.001 & 0.028 & 0.940 \\ 
& & 100 & -0.005 & 0.112 & 0.949 & -0.001 & 0.028 & 0.935 & -0.009 & 0.205 & 0.944 & 0.000 & 0.020 & 0.938 \\ 
  \multirow{-3}{*}{\centering 1} & \multirow{-3}{*}{\centering 30} & 200  & 0.000 & 0.081 & 0.944 & -0.001 & 0.022 & 0.902 & -0.009 & 0.146 & 0.946 & 0.000 & 0.015 & 0.923 \\ 
   \hline
 & & 30 & -0.010 & 0.269 & 0.939 & -0.004 & 0.105 & 0.903 & -0.015 & 0.391 & 0.950 & -0.003 & 0.079 & 0.933 \\ 
 & & 50 & -0.011 & 0.209 & 0.932 & -0.000 & 0.078 & 0.909 & -0.010 & 0.302 & 0.941 & 0.001 & 0.062 & 0.931 \\ 
& & 100 & 0.009 & 0.145 & 0.933 & -0.001 & 0.056 & 0.915 & -0.012 & 0.222 & 0.934 & -0.002 & 0.045 & 0.929 \\ 
  \multirow{-3}{*}{\centering 2} & \multirow{-3}{*}{\centering 10} & 200  & -0.008 & 0.105 & 0.931 & -0.002 & 0.038 & 0.934 & -0.007 & 0.150 & 0.960 & 0.001 & 0.031 & 0.935 \\ 
   \hline
 & & 30 & -0.005 & 0.229 & 0.943 & -0.001 & 0.079 & 0.930 & -0.014 & 0.377 & 0.951 & -0.002 & 0.067 & 0.940 \\ 
 & & 50 & -0.008 & 0.180 & 0.944 & 0.001 & 0.061 & 0.929 & -0.014 & 0.292 & 0.951 & -0.001 & 0.053 & 0.931 \\ 
& & 100 & 0.007 & 0.123 & 0.942 & 0.001 & 0.044 & 0.931 & -0.012 & 0.213 & 0.945 & -0.003 & 0.038 & 0.940 \\ 
  \multirow{-3}{*}{\centering 2} & \multirow{-3}{*}{\centering 20} & 200  & -0.007 & 0.090 & 0.933 & -0.001 & 0.030 & 0.939 & -0.006 & 0.147 & 0.957 & 0.001 & 0.026 & 0.945 \\ 
   \hline
 & & 30 & -0.006 & 0.216 & 0.943 & -0.001 & 0.070 & 0.939 & -0.014 & 0.374 & 0.951 & -0.002 & 0.062 & 0.946 \\ 
 & & 50 & -0.008 & 0.168 & 0.957 & 0.001 & 0.055 & 0.945 & -0.016 & 0.289 & 0.951 & -0.002 & 0.049 & 0.942 \\ 
& & 100 & 0.006 & 0.115 & 0.947 & -0.001 & 0.039 & 0.948 & -0.010 & 0.210 & 0.943 & -0.002 & 0.035 & 0.934 \\ 
  \multirow{-3}{*}{\centering 2} & \multirow{-3}{*}{\centering 30} & 200  & -0.004 & 0.084 & 0.935 & -0.000 & 0.027 & 0.940 & -0.008 & 0.145 & 0.950 & 0.000 & 0.025 & 0.942 \\ 
   \hline
 & & 30 & -0.048 & 0.245 & 0.949 & -0.043 & 0.075 & 0.725 & 0.048 & 0.341 & 0.951 & 0.057 & 0.060 & 0.629 \\ 
 & & 50 & -0.049 & 0.189 & 0.940 & -0.045 & 0.055 & 0.674 & 0.053 & 0.265 & 0.949 & 0.059 & 0.044 & 0.519 \\ 
& & 100 & -0.060 & 0.134 & 0.927 & -0.047 & 0.041 & 0.548 & 0.063 & 0.190 & 0.931 & 0.061 & 0.031 & 0.283 \\ 
  \multirow{-3}{*}{\centering 3} & \multirow{-3}{*}{\centering 10} & 200  & -0.052 & 0.095 & 0.907 & -0.046 & 0.029 & 0.355 & 0.064 & 0.135 & 0.924 & 0.061 & 0.022 & 0.079 \\ 
   \hline
 & & 30 & -0.029 & 0.216 & 0.945 & -0.024 & 0.051 & 0.798 & 0.016 & 0.351 & 0.955 & 0.028 & 0.038 & 0.766 \\ 
 & & 50 & -0.035 & 0.168 & 0.950 & -0.027 & 0.039 & 0.762 & 0.022 & 0.273 & 0.949 & 0.030 & 0.028 & 0.714 \\ 
& & 100 & -0.038 & 0.119 & 0.931 & -0.027 & 0.028 & 0.666 & 0.029 & 0.194 & 0.948 & 0.030 & 0.021 & 0.548 \\ 
  \multirow{-3}{*}{\centering 3} & \multirow{-3}{*}{\centering 20} & 200  & -0.034 & 0.083 & 0.935 & -0.027 & 0.019 & 0.514 & 0.031 & 0.137 & 0.953 & 0.031 & 0.014 & 0.272 \\ 
   \hline
 & & 30 & -0.023 & 0.207 & 0.946 & -0.017 & 0.041 & 0.847 & 0.005 & 0.354 & 0.954 & 0.018 & 0.031 & 0.832 \\ 
 & & 50 & -0.026 & 0.159 & 0.946 & -0.018 & 0.031 & 0.822 & 0.010 & 0.275 & 0.948 & 0.019 & 0.022 & 0.794 \\ 
& & 100 & -0.028 & 0.112 & 0.942 & -0.019 & 0.022 & 0.762 & 0.016 & 0.197 & 0.950 & 0.020 & 0.016 & 0.658 \\ 
  \multirow{-3}{*}{\centering 3} & \multirow{-3}{*}{\centering 30} & 200  & -0.024 & 0.079 & 0.941 & -0.019 & 0.015 & 0.628 & 0.018 & 0.139 & 0.950 & 0.021 & 0.011 & 0.438 \\ 
   \hline
\end{tabular}

\caption{Bias, standard deviation (sd) and coverage probability (cp) of 95\% nominal confidence interval for the fixed effect parameters in the simulation study. $n$ denotes sample size; $T$ denotes total number of observations for each individual; GM denotes generative model. The result is based on 1,000 replicates for each setting.}
\label{tab:simulation-result-full}
\end{table}

\begin{table}[htbp]
\centering

\begin{tabular}{rrrrrrrrrrrrr}
  \hline
& & & \multicolumn{2}{c}{$\sigma_{b2}^2$} & \multicolumn{2}{c}{$\sigma_{b3}^2$} & \multicolumn{2}{c}{$\sigma_{b0}^2$} & \multicolumn{2}{c}{$\sigma_{b1}^2$} & \multicolumn{2}{c}{$\sigma_\epsilon^2$} \\
\cmidrule(lr){4-5} \cmidrule(lr){6-7} \cmidrule(lr){8-9} \cmidrule(lr){10-11} \cmidrule(lr){12-13}
$n$ & $T$ & GM & bias & sd & bias & sd & bias & sd & bias & sd & bias & sd\\ 
  \hline
 & & 30 & 0.024 & 0.400 & - & - & -0.008 & 1.137 & - & - & -0.003 & 0.049 \\ 
 & & 50 & 0.013 & 0.300 & - & - & -0.020 & 0.868 & - & - & -0.002 & 0.035 \\ 
& & 100 & 0.017 & 0.210 & - & - & -0.031 & 0.614 & - & - & -0.001 & 0.024 \\ 
\multirow{-3}{*}{\centering 1} & \multirow{-3}{*}{\centering 10} & 200 & 0.004 & 0.151 & - & - & -0.021 & 0.431 & - & - & -0.000 & 0.017 \\ 
   \hline
 & & 30 & 0.012 & 0.319 & - & - & -0.025 & 1.067 & - & - & -0.003 & 0.032 \\ 
 & & 50 & 0.010 & 0.246 & - & - & -0.026 & 0.822 & - & - & -0.001 & 0.023 \\ 
& & 100 & 0.008 & 0.174 & - & - & -0.041 & 0.579 & - & - & -0.001 & 0.016 \\ 
\multirow{-3}{*}{\centering 1} & \multirow{-3}{*}{\centering 20} & 200 & 0.004 & 0.126 & - & - & -0.021 & 0.403 & - & - & -0.000 & 0.011 \\ 
   \hline
 & & 30 & 0.003 & 0.293 & - & - & -0.036 & 1.036 & - & - & -0.002 & 0.025 \\ 
 & & 50 & 0.001 & 0.232 & - & - & -0.037 & 0.809 & - & - & -0.001 & 0.018 \\ 
& & 100 & 0.008 & 0.163 & - & - & -0.040 & 0.569 & - & - & -0.001 & 0.013 \\ 
\multirow{-3}{*}{\centering 1} & \multirow{-3}{*}{\centering 30} & 200 & 0.000 & 0.116 & - & - & -0.023 & 0.399 & - & - & -0.000 & 0.009 \\ 
   \hline
 & & 30 & 0.047 & 0.498 & -0.001 & 0.058 & -0.003 & 1.238 & -0.003 & 0.040 & -0.003 & 0.048 \\ 
 & & 50 & 0.048 & 0.392 & -0.005 & 0.046 & -0.057 & 0.935 & -0.004 & 0.033 & -0.001 & 0.038 \\ 
& & 100 & 0.000 & 0.260 & -0.003 & 0.033 & -0.019 & 0.646 & -0.001 & 0.022 & 0.000 & 0.027 \\ 
\multirow{-3}{*}{\centering 2} & \multirow{-3}{*}{\centering 10} & 200 & 0.005 & 0.184 & -0.003 & 0.021 & -0.043 & 0.451 & -0.001 & 0.015 & -0.001 & 0.019 \\ 
   \hline
 & & 30 & 0.009 & 0.367 & -0.003 & 0.043 & -0.029 & 1.094 & -0.003 & 0.032 & -0.000 & 0.031 \\ 
 & & 50 & 0.022 & 0.302 & -0.003 & 0.033 & -0.045 & 0.854 & -0.002 & 0.025 & 0.000 & 0.025 \\ 
& & 100 & 0.002 & 0.200 & -0.002 & 0.021 & -0.016 & 0.597 & -0.000 & 0.017 & 0.001 & 0.017 \\ 
\multirow{-3}{*}{\centering 2} & \multirow{-3}{*}{\centering 20} & 200 & -0.001 & 0.142 & -0.001 & 0.015 & -0.029 & 0.418 & -0.001 & 0.012 & -0.001 & 0.012 \\ 
   \hline
 & & 30 & 0.001 & 0.334 & -0.002 & 0.036 & -0.045 & 1.065 & -0.003 & 0.029 & 0.000 & 0.025 \\ 
 & & 50 & 0.012 & 0.268 & -0.003 & 0.027 & -0.049 & 0.826 & -0.002 & 0.022 & 0.000 & 0.019 \\ 
& & 100 & 0.002 & 0.183 & -0.001 & 0.019 & -0.028 & 0.584 & 0.000 & 0.016 & 0.000 & 0.013 \\ 
\multirow{-3}{*}{\centering 2} & \multirow{-3}{*}{\centering 30} & 200 & -0.003 & 0.127 & -0.001 & 0.013 & -0.029 & 0.409 & -0.001 & 0.011 & -0.000 & 0.009 \\ 
   \hline
 & & 30 & 0.126 & 0.434 & - & - & -0.710 & 1.159 & - & - & 0.004 & 0.046 \\ 
 & & 50 & 0.105 & 0.329 & - & - & -0.771 & 0.860 & - & - & 0.005 & 0.034 \\ 
& & 100 & 0.094 & 0.228 & - & - & -0.810 & 0.604 & - & - & 0.005 & 0.025 \\ 
\multirow{-3}{*}{\centering 3} & \multirow{-3}{*}{\centering 10} & 200 & 0.080 & 0.159 & - & - & -0.796 & 0.429 & - & - & 0.006 & 0.018 \\ 
   \hline
 & & 30 & 0.059 & 0.329 & - & - & -0.380 & 1.056 & - & - & 0.000 & 0.029 \\ 
 & & 50 & 0.053 & 0.262 & - & - & -0.428 & 0.800 & - & - & 0.001 & 0.023 \\ 
& & 100 & 0.040 & 0.174 & - & - & -0.429 & 0.575 & - & - & 0.001 & 0.017 \\ 
\multirow{-3}{*}{\centering 3} & \multirow{-3}{*}{\centering 20} & 200 & 0.038 & 0.125 & - & - & -0.430 & 0.406 & - & - & 0.002 & 0.011 \\ 
   \hline
 & & 30 & 0.040 & 0.304 & - & - & -0.268 & 1.029 & - & - & -0.000 & 0.024 \\ 
 & & 50 & 0.030 & 0.237 & - & - & -0.296 & 0.782 & - & - & -0.001 & 0.018 \\ 
& & 100 & 0.027 & 0.162 & - & - & -0.306 & 0.569 & - & - & 0.000 & 0.013 \\ 
\multirow{-3}{*}{\centering 3} & \multirow{-3}{*}{\centering 30} & 200 & 0.023 & 0.115 & - & - & -0.299 & 0.395 & - & - & 0.001 & 0.009 \\ 
   \hline
\end{tabular}

\caption{Bias and standard deviation (sd) for the estimated variance components $\sigma_{bj}^2,0 \leq j \leq 3$ and $\sigma_\epsilon^2$ in the simulation study. $n$ denotes sample size; $T$ denotes total number of observations for each individual; GM denotes generative model. For GM 1 and GM 3, the model doesn't include $b_{i1}$ and $b_{i3}$, so the corresponding entries in the table are left blank. The result is based on 1,000 replicates for each setting.}
\label{tab:simulation-result-varcomp}
\end{table}

\end{landscape}
}

\section*{Acknowledgements}

Research presented in this paper was supported by the National Heart, Lung and Blood Institute under award number R01HL125440; the National Institute on Alcohol Abuse and Alcoholism under award number R01AA023187; the National Institute on Drug Abuse under award number P50DA039838; and the National Institute of Biomedical Imaging and Bioengineering under award number U54EB020404.
The authors would like to thank Peng Liao, Walter Dempsey, two reviewers, and the associate editor for helpful suggestions.

\bibliographystyle{imsart-nameyear}
\bibliography{lmm-ref}

\end{document}